\documentclass[a4paper]{spie}  
\usepackage{graphicx}
\usepackage{subcaption}
\usepackage{multirow}
\usepackage{amsmath,amsfonts,amssymb}
\usepackage{graphicx}
\usepackage[colorlinks=true, allcolors=blue]{hyperref}

\title{FiberPol-6D: Spectropolarimetric Integral Field mode for the SAAO 1.9~m Telescope using fibers}

\author[a]{Siddharth Maharana}
\author[a,b]{Sabyasachi Chattopadhyay}
\author[a,c]{Matthew Bershady}
\affil[a]{South African Astronomical Observatory, 1 Observatory Rd., 7935, Cape Town, South Africa}
\affil[b]{Centre for Space Research, North-West University, Potchefstroom 2520, South Africa}
\affil[c]{University of Wisconsin–Madison, 475 N. Charter St., Madison, WI, USA}

\authorinfo{Further author information: (Send correspondence to S.M.)\\S.M.: E-mail: siddharth@saao.ac.za, sidh345@gmail.com}
\begin{document} 
\maketitle

\begin{abstract}
\par Most optical spectropolarimeters built to date operate as long-slit or point-source instruments; they are inefficient for observations of extended objects such as galaxies and nebulae. 2D spectropolarimetry technique development is a major challenge in astronomical instrumentation. At the South African Astronomical Observatory’s (SAAO) FiberLab, we are developing a spectropolarimetry capable Integral Field front-end called FiberPol(-6D) for the existing SpUpNIC spectrograph on the SAAO’s 1.9 m telescope. SpUpNIC is a general purpose 2 arc-minute long-slit spectrograph with a grating suite covering the wavelength range from 350 to 1000~nm and at spectral resolutions between 500 and 6000. FiberPol generates 6D observational data: x-y spatial dimensions, wavelength, and the three linear Stokes parameters $I$, $q$ and $u$. Using a rotating half-wave plate and a Wollaston prism, FiberPol executes two-channel polarimetry, and each channel is fed to an array of 14 fibers, corresponding to a field of view of $10\times20~{\rm arcseconds}^{2}$ sampled with 2.9 arcsec diameter fiber cores. These fiber arrays are then rerouted to form a pseudo-slit input to SpUpNIC. FiberPol aims to achieve a polarimetric accuracy of 0.1\% per spectral resolution bin. Further, it can also function as a non-polarimetric integral-field unit of size $20\times20~{\rm arcseconds}^{2}$. 

\par The instrument design has been completed and it is currently being assembled and characterized in the lab.  It is scheduled for on-sky commissioning in the second half of 2024.  In this paper, we present the scientific and technical goals of FiberPol, its overall design and initial results from the lab assembly and testing. 

\par FiberPol is a low-cost technology demonstrator ($< 10,000$ USD), and the entire  system predominantly employs small size (one inch or less), commercial off-the-shelve optics and optomechanical components. It can be modified and replicated for use on any existing spectrograph, especially on bigger telescopes like the 10~m Southern African Large Telescope (SALT) and the upcoming 30~m class telescopes. 

\end{abstract}


\section{INTRODUCTION}
\label{sec:intro}  

Polarimetry refers to the measurement of the polarization state of light coming from a source. It entails ascertaining the amount and direction/sense of any preferred polarization state in the light beam. Polarimetry is a powerful tool that has been used by astronomers to understand the physics of diverse classes of astrophysical objects, such as active galactic nuclei (AGN), supernovae, protoplanetary systems, and dust clouds in the interstellar medium (ISM) \cite{Hough_review,Scarrott-1991}. In particular, it is useful in the study of objects that have an inherent asymmetry in their light emission, absorption or propagation mechanism. Often, polarimetry is the only method to find the geometry of an astrophysical system and cannot be obtained by other methods such as imaging and spectroscopy. 

\par Astronomers have been building polarimeters, with ever-increasing precision and accuracy with most present-day instruments reaching accuracies of the order of 0.1\% or better \cite{hippi2, DIPOL2, robopol} in the optical wavelengths. Most modern optical polarimeters in astronomy come in two broad kinds: imaging polarimeters and spectropolarimeters. As the name suggests, imaging polarimeters create a polarization (Stokes parameters) map of a sky field in different broadband filters, while spectropolarimeters enable polarization measurement as a function of wavelength. In general, spectropolarimeters are a more powerful tool to probe the physics of astronomical sources, just as the spectra of a source has more encoded information than only the intensity information obtained through images. Spectropolarimetry has enabled astronomers in the past to diagnose very difficult astrophysical problems such as the physics of flaring and jetted compact objects. A seminal example is the studies that led to the unification of the Type I and II AGNs into one class\cite{agn_unification}.

\par Due to limitations of available instrumentation technology, spectropolarimetry has till date been available only in long-slit mode, effectively allowing for observations of only point sources. Thus, development of polarimetry-capable Integral Field Units (IFUs) will enable measurements of extended objects, hitherto unfeasible with current day instruments. Such a system, generically referred to as a Integral Field Spectro-polarimeter (IFSP), will generate 6D data as output- two spatial dimensions $x, ~y$, wavelength $\lambda$, intensity $~I,$ and Stokes parameters $ ~q$, and ~$u$ (or $p$ and $\theta$)\footnote{If the circular Stokes parameter $v$ is included, it will be a 7D data system.}. 

\par While polarimetry (including spectropolarimetry) and IFUs \cite{bershady20093d} have become a commonly available observational tool for most astronomers, IFSP has rarely been attempted. We refer interested readers to an excellent introductory paper on IFSP concepts and possible instrument design ideas by Rodenhuis et al.\cite{Snik_keller_IFSP}. To tackle the challenge of building a simple and easily adaptable IFSP system for astronomy, we are developing FiberPol(-6D), a fiber-fed IFSP system. FiberPol will be deployed as a front-end system on the existing SpUpNIC spectrograph\cite{SpUpNIC} on the South African Astronomical Observatory's (SAAO) 1.9 m Radcliffe telescope at Sutherland. SpUpNIC is a general purpose 2 arc-minute long-slit spectrograph with a grating suite covering the wavelength range from 350 to 1000 nm and at spectral resolutions between 500 and 6000. 

\par The primary science objective with FiberPol is to study the dust properties in the various astrophysical environments including the ISM of nearby galaxies, and objects within the Milky Way such as nebulae, molecular clouds, supernova remnants, asteroids and comets. Even for point objects such as supernovae and AGN, the 2D field will be efficient in measuring the ISM introduced foreground polarization by simultaneous observation of field stars, thus allowing astronomers to accurately estimate the inherent polarization of the source. Spectropolarimetry is an invaluable tool to study the interstellar dust\cite{andersson_review}. It is used most often to measure the wavelength dependent polarization of starlight due to the dichroic extinction by interstellar dust aligned with the ambient magnetic field. The shape of the polarization distribution over wavelength is very well constrained by the ``Serkowski function"\cite{Serkowski1975}, and is ubiquitously found for all sight-lines in the Milky Way. The Serkowski function parameters place direct constraints on the size, nature and alignment of dust grains, and is one of the key ingredients of modern dust models \cite{Hensley_2023}. While Serkowski curves for sight-lines in the Milky Way have been examined with great detail, similar studies have not been done for other galaxies. External galaxies, being extended objects would need to be observed through an IFSP mode. Only with such observations, we will be able to test and extend our understanding of current dust models and grain alignment theories to the local Universe.

\par In addition to developing a general purpose fiber-based IFSP technology, a coequal motivation for building FiberPol is to demonstrate high-accuracy optical polarimetry using fibers coupled to a general spectrograph. This demonstration opens a path to relatively inexpensive and retrofittable polarimetric instruments accessible to the astronomy community on telescopes of all sizes. Currently, polarimetry is done through specialized and dedicated instruments (polarimeters), which are very expensive and difficult to maintain. As a result, polarimetry is often relinquished on many small, and almost all large telescopes due to the requirements of large, expensive and fragile optics, and the challenging optical designs to accommodate these. FiberPol opens up the possibility of using relatively small, inexpensive fiber-fed polarimetric units to pre-process the polarization signal with very little light loss. In the past, there have been successful efforts in using fibers with polarimetric instrumentation, especially to relay polarimetrically analyzed light beams to bench spectrographs for high-resolution spectropolarimetry measurements\cite{HARPSPol, PEPSI}. We capitalize on this fibre-based pre-processing concept and extend it to integral field spectroscopy.

\par In this paper, we present the overall design of FiberPol and its current development status. Section~\ref{tech_goals_sec} introduces the top-level concept and technical goals for FiberPol. The optical and optomechanical design of the instrument is described in Section~\ref{design_sec}. The current status of the instrument, the next steps in the development and schedule for the commissioning is presented in Section~\ref{colculsions_sec}. 

\section{FiberPol concept: technical goals and constraints}
\label{tech_goals_sec}

\begin{table}[]
\centering
\caption{FiberPol host -- details of the SAAO 1.9 m telescope and SpUpNIC spectrogaph}
\begin{tabular}{ccc}
\hline
Sl. No. & Parameter                & Value                                   \\ \hline
1 & Telescope f/\#           & 18.0                                    \\
2 & Telescope Aperture       & 1.88 m                                  \\
3 & Slit length              & 2’ ($\sim$37 mm)                        \\
4 & Median seeing full-width half-maximum  & 1.5"\\
5 & Observatory ambient temperature        & $-10^{\circ}$ to $20^{\circ}$\\
6 & Telescope Port           & Direct Cassegrain Port                  \\
7 & SpUpNIC Wavelength Range & 350 nm to 1000 nm                       \\
8 & Spectral Resolution      & 500 to 6500                             \\
9 & No of gratings           & 10                                      \\ \hline
\end{tabular}
\label{SpUpNIC_spec}
\end{table}

\par
\begin{figure}[h!]
    \centering
    \includegraphics[scale=0.225]{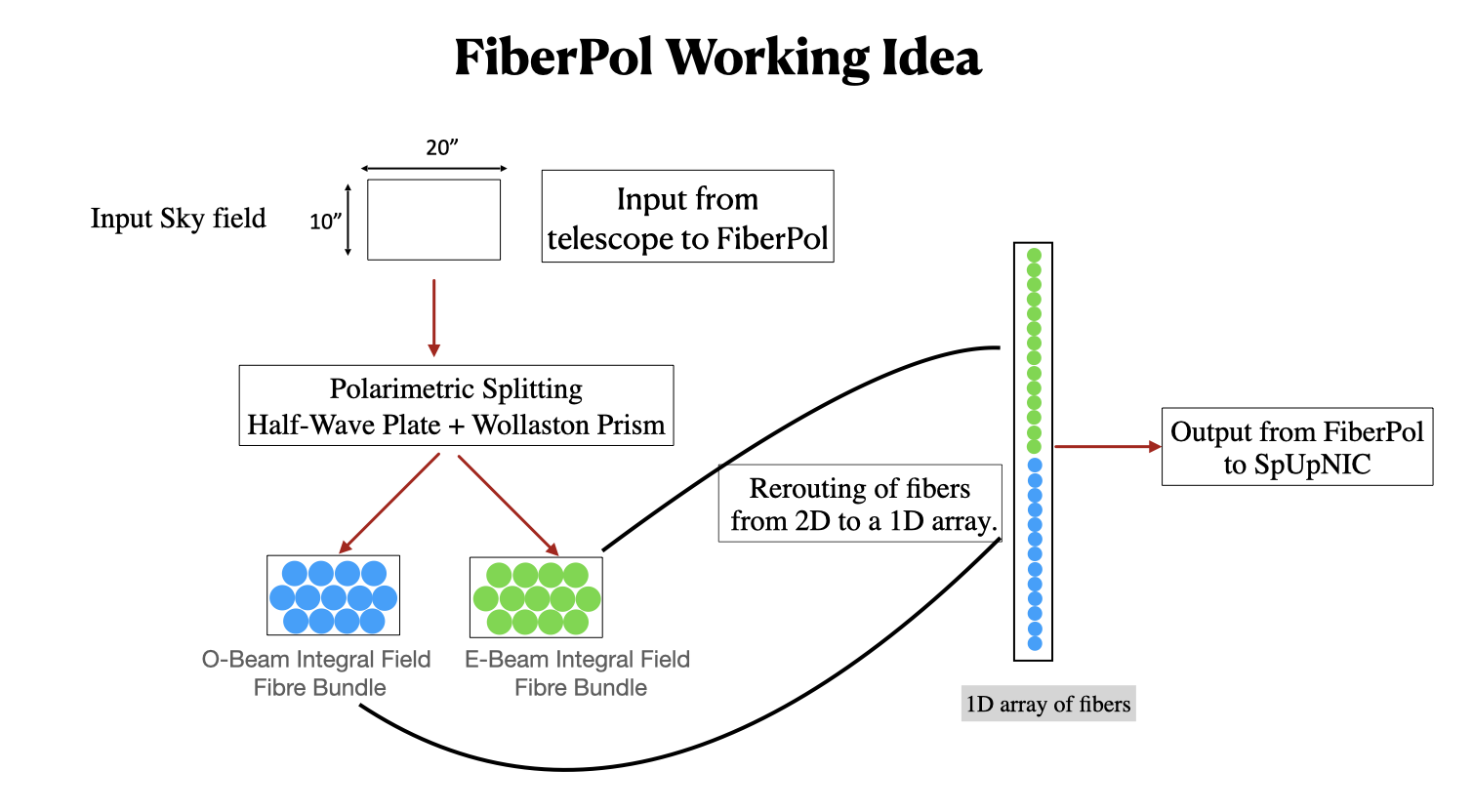}
    \caption{\footnotesize{FiberPol top-level concept: light from a 2D object on sky, in the shape of a fiber-array of size of roughly 10$\times$20~sq. arcsec passes through a polarization analyzer system, which splits the light into orthogonal polarization components (called $e$ and $o$ beams). Each beam is fed to an array of 14 fibers. These fiber arrays are rerouted to form a pseudo-slit input to SpUpNIC, such that the input to the spectrograph remains a 1D array/object at the nominal telescope $f/\#$. As output from SpUpNIC, we will get 28 spectra for the 14 spatial positions of the 2D object. For each spatial position, the normalized difference between the $e$ and $o$ spectra will yield the spectro-polarimetric data.}}
    \label{FiberPol_concept}
\end{figure}

 The schematic of the top-level concept of FiberPol is shown in Figure~\ref{FiberPol_concept}. Using a rotating half-wave plate and a Wollaston prism (WP) as the polarization analyzer system, FiberPol executes two-channel linear polarimetry, and each channel is fed to an array of 14 (13 object and 1 sky) fibers, corresponding to a field of view (FoV) of $10\times20~{\rm arcseconds}^2$. These fiber arrays are then rerouted to form a pseudo-slit input to SpUpNIC, such that the input to the spectrograph remains a 1D array/object. As output from SpUpNIC, we obtain 28 spectra, one for each fiber. For each of the 14 spatial positions on sky, the normalized difference between the corresponding spectra from each channel (referred to as $e$ and $o$ spectra) yields the Stokes parameter values across wavelength. In addition to a polarimetric mode, we have made the provision for a non-polarmetric and conventional IFU mode with FiberPol. This IFU system will have twice the FoV area compared to the polarimetric mode; details are given in Section~\ref{reducer_optical_design_sec}. We design FiberPol to function as an add-on mode that can be easily deployed, on-demand, before the night begins. Thus, FiberPol will introduce IFSP capability to SpUpNIC without making any changes to the telescope or instrument. The specifications of the telescope, SpUpNIC and the site conditions are listed in  Table~\ref{SpUpNIC_spec}; FiberPol has been designed accordingly.


\subsection{Why a 2-channel design}

\par The technical goals of the FibePol instrument are captured in Table~\ref{tech_goals}. Polarization signals at optical wavelengths, especially emanating from dichroic absorption and scattering by Galactic dust are very small, often at levels of 0.5~\% or less, requiring polarimeters to achieve accuracies of 0.1~\% or better\cite{Skalidis}. Obtaining such high accuracies makes calibration of polarimeters a challenging instrumentation problem. With FiberPol, the critical design goal is to achieve a polarimetric accuracy of 0.1\% per spectral and spatial resolution bin. Two-channel or four-channel polarimetry has been demonstrated as the most accurate way of measuring polarization. Astronomers very often use a WP as the main analyzer element instead of a simple linear polarizer due to the following reasons: (1) While the sky is not birefringent at optical wavelengths, atmospheric conditions such as transparency can change during the course of observations. This can lead to erroneous polarization measurements unless  the two intensity measurements for the $q$ and $u$ parameters are made simultaneously. This is mitigated by using WP like beamsplitters which separate the two orthogonal polarization states, called the ordinary ($o$) and extraordinary ($e$) beams, and allow their intensities to be measured together. (2) By using a simple polarizer, for low polarization sources encountered in astronomy, half the amount of light received from the source (whose polarization is orthogonal to the polarizer axis) is lost. The standard practise is to place a WP at a pupil plane inside the instrument and the incident pupil is split symmetrically along the optical axis into $o$ and $e$ beams. At the detector plane, each source/star appears as two images, called the $o$ and $e$ images.

\par  In FiberPol, \textit{the light from the objects is fed to fibers only after the polarimetric analysis by the WP has been executed}. This ensures that polarization scrambling behaviour of fibers is mitigated. From thereon, the only effect that fibers and downstream optics can introduce is of preferential transmission of one beam (polarization) over the other; this is a commonly encountered effect in all polarimeters and will need to be corrected by careful polarimetric calibration using on-sky standard sources. While polarimetric calibration methods for narrow and wide FoV polarimeters exist\cite{RoboPol_standards, WALOP_Calibration_paper, sky_pol}, as IFSP has not been attempted before, we will develop methods to carry out accurate calibration for FiberPol.

\begin{table}[]
    \centering
    \caption{FiberPol system design goals}    \begin{tabular}{ccc}
        \hline
        \textbf{Sl. No}. & \textbf{Parameter} & \textbf{Technical Goal} \\
        \hline
        1 & Polarimetric Accuracy  & 0.1~\%\\
        2 & Polarimeter Type & Two Channel Linear Polarimetry \\
        3 & Polarization Analyzer Type & Wollaston Prism and a rotating half-wave plate \\
        4 & Field of View & $10"\times20"$\\
        5 & PSF Performance & Seeing limited at fiber-injection plane\\
        6 & f/\# at fiber feed  & 3.5 to 4.5 \\
        7 & Telecentricity at fiber-injection plane & $=<~0.1^{\circ}$ \\
        8 & No of fibers per beam & 13 object \& 1 sky \\
        9 & Optimized Wavelength Range & 400$\mu$m - 700$\mu$m\\
        10 & Optics and Optomechanical component & Off-the-shelve \\
        11 & Cost & Under $10k$~USD\\
        12 & Design flexibility & Design concept should work for any long-slit spectrograph.\\
        13 & Fiber Aperture & $100/140~{\mu}m$ core/cladding \\
        \hline
    \end{tabular}
    \label{tech_goals}
\end{table}

\subsection{Beam speed considerations}

\par The telescope nominally feeds a very slow f/18 beam as input to the long-slit of SpUpNIC. Fibers induce significant focal ration degradation (FRD) at such slow speeds. To minimize light-loss due to FRD we aimed to feed the fibers at $f/\#$ between 3.5 to 4.5. While FRD will still introduce a faster fiber output, the change is modest. The best-fit to laboratory data on fiber FRD yields the following equation:
\begin{equation}\label{frd}
    f_{o} = 2.6998\times ln(f_{i}) - 0.08
\end{equation}  
where $f_o$ is the output $f/\#$ corresponding to 98~\% encircled energy for a given input  $f/\# = f_i$ (see Chattopadhyay et al. \cite{FRD_Sabyasachi_paper} for a detailed exposition on this subject). We use this equation to design a reimaging system that captures up to 98\% of the light from the telescope.

\par To account for fiber FRD, in addition to separating the orthogonal polarizations using a WP, before feeding the beams to the fibers, the $e$ and $o$ beams need to be sped up to a faster beam. Both these tasks are done by the subsystem called ``Reducer" (i.e., a focal-reducer), which is described later. The input $f/\#$  to the fibers is 3.7 and based on Equation~\ref{frd}, the output is expected to be 3.45. The other main subsystem of the instrument, called ``Expander" (i.e., a focal expander), takes as input the fibers arranged in from of a 2' pseudoslit at $f/3.45$, and expands the beam back to $f/18$ to feed SpUpNIC. A contributing factor to enhanced FRD is non-telecentricity in the input beam to the fibers, as discussed by Chattopadhyay at al., where they detail the importance of feeding telecentric beams. We aim to restrict the non-telecentricity to $<0.1^{\circ}$ for all wavelengths and field points.

\subsection{Available field}

\par The slit length of 2' is a physical limit of the SpUpNIC spectrograph; this places a constraint on the total number of fibers in the $e$ and $o$ arrays each. In the first phase of the project, we are designing IFUs with $100~{\mu}m$ core fibers as their mounts can be accurately fabricated by the SAAO workshop with existing wire-EDM tooling. These cores sample 2.93 arcsec in diameter on sky. In the subsequent phase of the project, we plan to make IFUs using $50~{\mu}m$ core ($85~{\mu}m$ buffer) fibers, which will allow more spatial points on sky with finer spatial sampling, at the expense of smaller FoV and lower fill factor. In our case, the total number of allowable $100~\mu m$ core fibers with a $140~\mu m$ outer diameter (cladding and buffer) comes out to be 28 at the $f/3.45$ focal plane. 

\subsection{Cost considerations}

\par As one of the main goals of FiberPol is to demonstrate an affordable IFSP technology using fibers and small-size and low-cost optics, we decided to design FiberPol employing commercially available, off-the-shelve optical and opto-mechanical components. Thus, except for the fiber-mounts, all optics and opto-mechanical components have been designed from existing vendor catalogs. The optimized wavelength range for FiberPol system is 400 to 700~nm for two congruent reasons: (a) anti-reflection coatings optimized for this wavelength window are available for most off-the-shelve optical components, and (b) the peak of the Serkowski curve is usually found to be centered around 550~nm. FiberPol is designed as a front-end system for the existing spectrograph, SpUpNIC; as such, FiberPol must be compact and fit in a small volume inside the acquisition box sitting above the spectrograph. This allowable volume spans roughly total 18~cm in length along the optical axis with more range in the other two dimensions. While conforming to the above constraints of volume, coatings and lens prescriptions, FiberPol needs to deliver high-quality (seeing limited) imaging and throughput performance ($\sim80\%$, excluding telescope and SpUpNIC losses). 



\section{FiberPol Design}\label{design_sec}

\subsection{Choice of Polarizer System}

\begin{table}[h!]
    \centering
    \caption{Details of FiberPol polarizer optics}
\begin{tabular}{ccc}
    \hline
       Part & Part Description & Vendor \& Part\# \\
        \hline
         Wollaston Prism &  $MgF_{2}$ ~$45^{\circ}$ wedges, 11\time11~mm aperture & Thorlabs (WPM10)\\
         Half-wave plate & Quartz and $MgF_{2}$ plates, one inch aperture & Thorlabs (AHWP10M-600)\\
         \hline
    \end{tabular}
    \label{pol_optics_table}
\end{table}

\par After a careful study of multiple potential optical designs, we completed the instrument's optical and subsequent optomechanical system design. The instrument's polarization analyzer subsystem was at the core of the optical design. We considered the following three options. 

\begin{enumerate}
    \item Savart Plates. Owing to the large birefringence of calcite, a calcite plate, when placed near the focal plane, can split the beam into orthogonal polarization components with linear displacement between them. While this is an attractive feature, calcite is unsuitable for use in broad wavelength applications due to the strong birefringence dependence on wavelength, causing the beams coming out of the plates to be dispersed. This effect has been explained in detail in the optical design paper of the Wide-Area Linear Optical Polarimeter (WALOP)-South instrument that employed calcite Wollaston prisms\cite{WALOP_South_Optical_Design_Paper}. While other birefringent materials such as quartz and magnesium fluoride ($MgF_{2}$) have negligible birefringence dependence on wavelength, their birefringence value is an order of magnitude smaller than calcite, and would require extremely thick ($> \rm{10~cm}$)and custom made plates.
    
    \item Polarization Beam-Splitter (PBS): a wire-grid or dielectric coated PBS splits an incident beam into orthogonal polarization states at $90^{\circ}$. Wire-grid PBS have found application in two-channel polarimeters such as MOPTOP\cite{moptop}, where using a rotating half-wave plate (HWP) as the polarization modulator, a moderate on-sky accuracy of 0.25~\% was achieved. A key advantage for the PBS is the near-constant beam-splitting performance over large angles of incidence and broad wavelength ranges. A critical drawback, especially in case of instruments that aim high accuracy polarimetry is that although the transmitted beam has a high extinction ratio of $>~10^{4}$, the corresponding value for the reflected beam is less than 100.
    
    \item WP: A WP with a modulating HWP system has been used in astronomy as the default analyzer system whenever aiming for accuracies $>=0.1\%$. WPs separate the orthogonal polarization states (called $e$ and $o$ beams) with high extinction ratios ($>~10^{4}$). When the beam splitting by the WP is done at the pupil plane, each source/star appears as two ($e$ and $o$) images at the detector plane, at almost equal distance from the nominal position of the star/source.
\end{enumerate}

 Usually, WPs (and other polarimetric beam-splitters) split $0^\circ$ and $90^\circ$ polarizations from which $q$ can be measured. In order to measure $u$, instead of rotating the whole instrument along the optical axis in the Instrument Coordinate System (ICS), the HWP is used to rotate the plane of polarization of the incoming beam with respect to the ICS. A HWP, with its fast axis at an angle $\alpha$ with the ICS rotates the polarization plane by angle $2\alpha$. A typical polarization measurement cycle for a two channel polarimeter consists of measurements at HWP angles of $0^\circ$, $22.5^\circ$, $45^\circ$ and $67.5^\circ$: while the first two correspond to $q$ and $u$ measurements, the latter two correspond to nominal measurements of $-q$ and $-u$ (which enables quantification and correction of instrument induced polarization, called \textit{instrumental polarization}).
Due to the ability for high-accuracy polarimetry and symmetrical splitting of the beams, we chose a WP+HWP as the polarization analyzer system for FiberPol. The following subsections describe the optical design of FiberPol.

\begin{figure}[h!]
    \centering
    \includegraphics[scale=0.325]{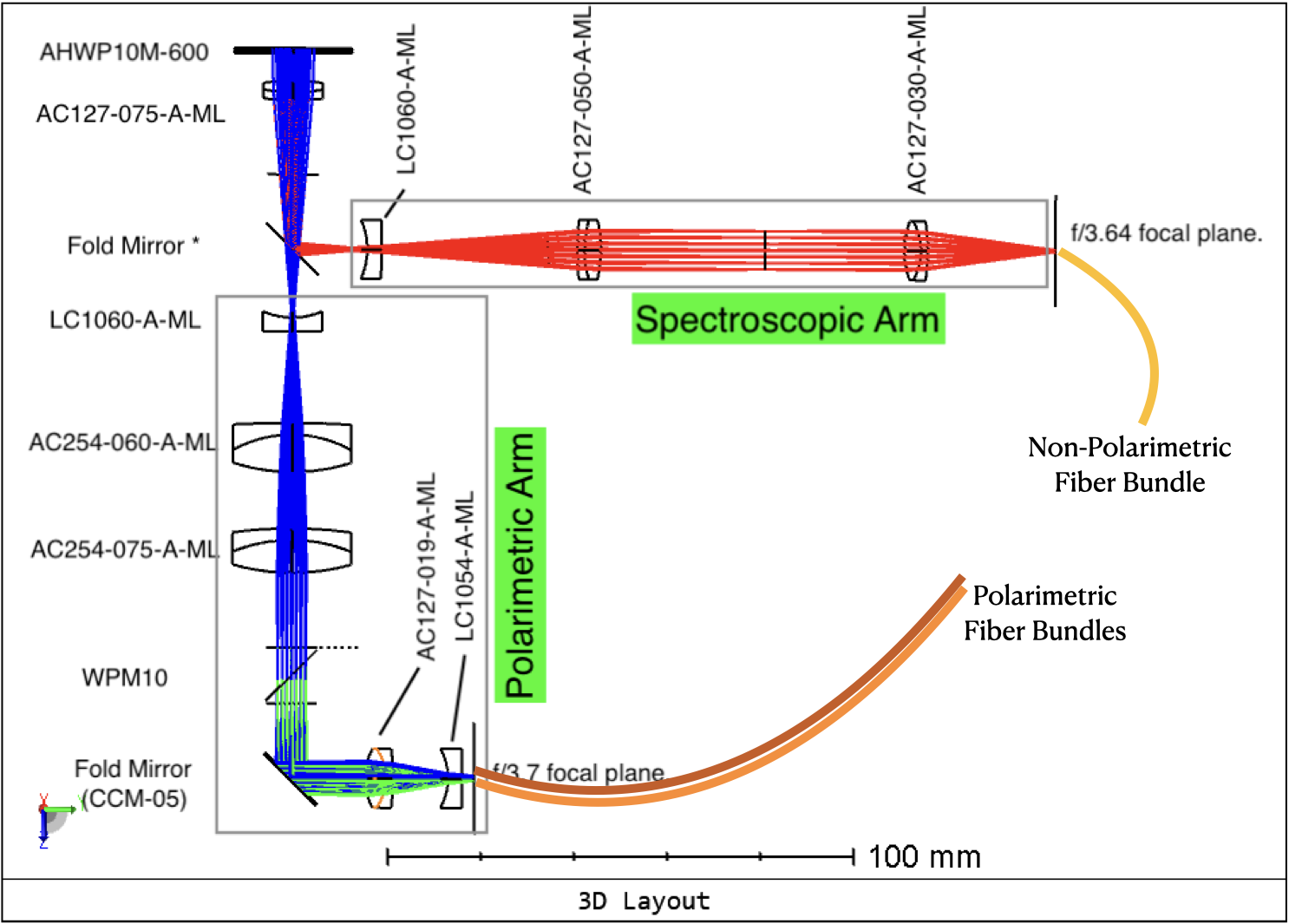}
    \caption{Optical design of the FiberPol Reducer system. The polarimetric and non-polarimetric arms of FiberPol are overlaid in the figure. Light from the telescope comes from the top and it passes through a half-wave plate and then sent to polarimetric or spectroscopic arm using the retractable fold mirror. All the optical components have been selected from the Thorlabs catalog, with part names annotated in the image.}
    \label{reducer_optical_design}
\end{figure}

\subsection{Reducer Optical Design}
\label{reducer_optical_design_sec}

\par Figure~\ref{reducer_optical_design} shows the optical design layout of the FiberPol Reducer system. A half-wave plate (HWP), mounted on motorized rotary stage is the first element, followed immediately by a half inch achromatic field-reducer lens that creates an intermediate $f/6.8$ focus. Just before this focus, there is a provision for a fold mirror to be inserted in the path when the non-polarimetric (spectroscopic) mode is used. After the intermediate focus, for the polarimetry mode, a set of three lenses of sizes half and one inch create a collimated beam of size about 9~mm. At the pupil, a $MgF_{2}$ WP is placed at the pupil. The WP splits the $e$ and $o$ beams symmetrically at $1.2^{\circ}$ at 500~nm. The collimated $e$ and $o$ beams pass through a fold mirror, after which two lenses create a flat and telecentric ($=<~0.1^{\circ}$ across the FoV) $f/3.7$ focal plane for the fiber feed. The fold mirror is placed after the WP to make the design compact. The HWP is 1 inch in aperture, made of quartz and $MgF_{2}$ plates that has been designed for achromatic performance in the wavelength range of 400 to 800~nm. Table~\ref{pol_optics_table} listed the details and source of the polarizer optics used in the FiberPol design. The design employs two telecentricity-correcting concave field lens, one each after the $f/6.8$ and before the final $f/3.7$ focal plane. In addition to allowing for a gradual step-down of the $f/\#$ of the beam from $f/18$ to $f/3.7$, the $f/6.8$ intermediate focus provides a plane where a physical mask (field stop) is placed to avoid $e$ polarized light from outside the FoV entering the $o$-beam fiber array footprint at the final $f/3.7$ focal plane, and vice-versa. For the spectroscopic arm, the design consists of a telecentricity correcting concave field lens followed by a combination of collimator and focusing half inch achromatic lenses.

\par The major challenges in creating the optical design were to meet the goals of obtaining seeing limited PSFs at the $f/3.7$ focal-plane while achieving better than $0.1^{\circ}$ non-telecentricity in the constrained instrument volume. We minimized the number of optical elements in the design to incur minimal reflection losses.  All the lenses used in the design are available as pre-mounted lenses that can then be easily integrated in a cage system. This makes the assembly and alignment process easier, as we do not require high-precision optical alignment, as described later.

\par
\begin{figure}
    \centering
    \includegraphics[scale=0.3375]{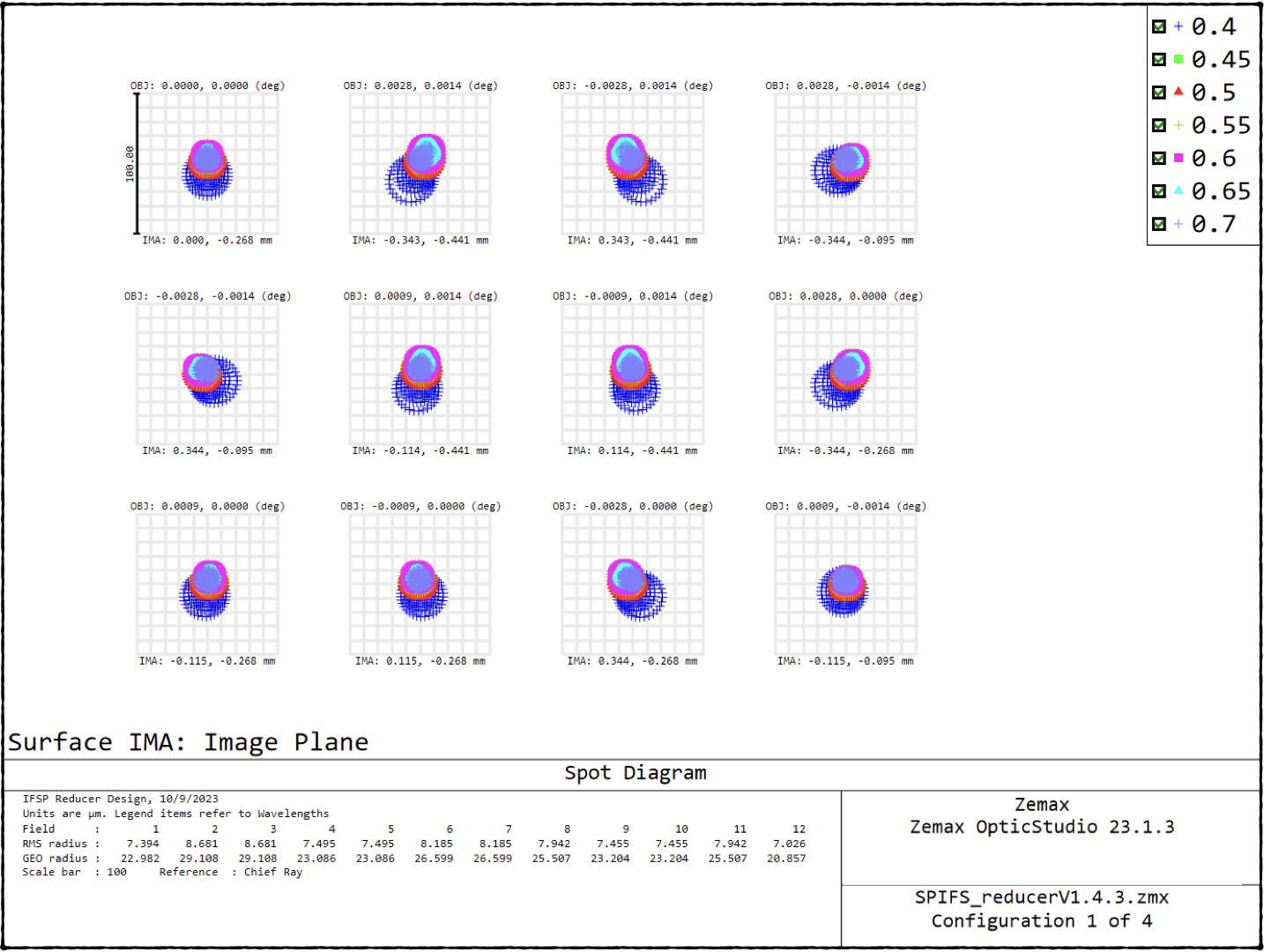}
    \caption{Spot diagram for one of the two polarized beams at the  $f/3.7$ fiber-feed focal plane. Colors, labelled in the legend, represent the wavelengths for which FiberPol is optimized. RMS and GEO radii are the root-mean square and geometric radii of the spot diagrams, respectively. Spot diagrams for the $o$ and $e$ beams are nearly identical as they have similar optical paths. For reference, the RMS corresponding to the median seeing of 1.5" is 19.2 ${\mu}m$. The RMS and GEO radii for the spectroscopic arm are smaller.}
    \label{reducer_spot_diagram}
\end{figure}

 Figure~\ref{reducer_spot_diagram} shows the spot diagram for the FoV of the polarimetric arm for one of the two beams. The spot diagrams for the $o$ and $e$ beams are nearly identical as they follow similar optical paths. For reference, the root-mean-square (RMS) spot size corresponding to the median seeing of 1.5" is 19.2 ${\mu}m$. Thus with the presented optical design, the spot diagram RMS radii of $\sim7~{\mu}m$ is around 3 times better the seeing RMS, which will enable seeing limited PSFs if the alignment of the optics is maintained in the assembly. To estimate the effect of assembly misalignment on the optical performance, we carried out a tolerance analysis of the system. The summary of the analysis is the following: if the alignments can be maintained within $\pm 50~{\mu}m$ for air-gap separations between lenses and $\pm 100~{\mu}m$ for centering and tilt of $\pm 2'$ of individual lenses, the degraded RMS radii will be under 8${\mu}m$ for the full FoV.



\par
\begin{figure}
\centering
\begin{subfigure}[b]{0.69\textwidth}
    \centering
    \includegraphics[scale=0.35]{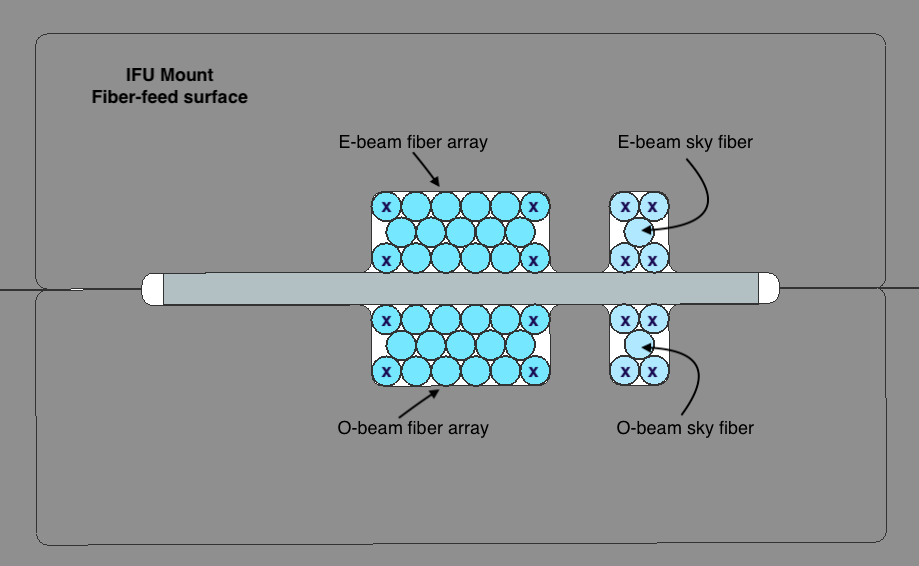}
    \label{ifu_mount}
\end{subfigure}
\begin{subfigure}[b]{0.3\textwidth}
    \centering
    \includegraphics[scale=0.21]{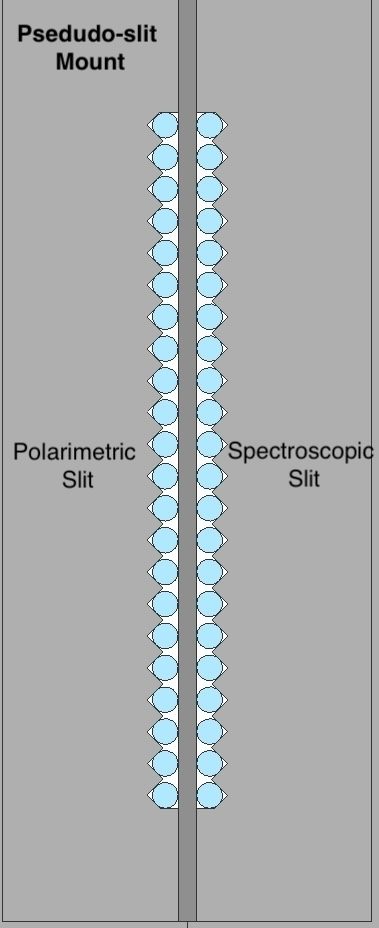}
    \label{slit_mount}
\end{subfigure}
\caption{CAD model of the polarimetric IFU (left) and pseudo-slit (right) ends of the fiber arrays. The cross marked fibers are dummy fibers present to ensure accurate packing. The pseudo-slit for the polarimetric and spectroscopic arms are placed side by side and fed to the Expander system.}
\label{IFU_slit_fig}
\end{figure}

 The nominal separation between the $e$ and $o$ images at the $f/3.7$ focal plane is 0.536 mm. This separation allows sufficient space for three rows of fibers to be placed for each channel with a 150~${\mu}m$ separation between them (and in turn drives the geometry of the fiber arrays), as shown in Figure~\ref{IFU_slit_fig}(a).

\subsection{Expander Optical Design}

\begin{figure}
    \centering
    \includegraphics[scale=0.25]{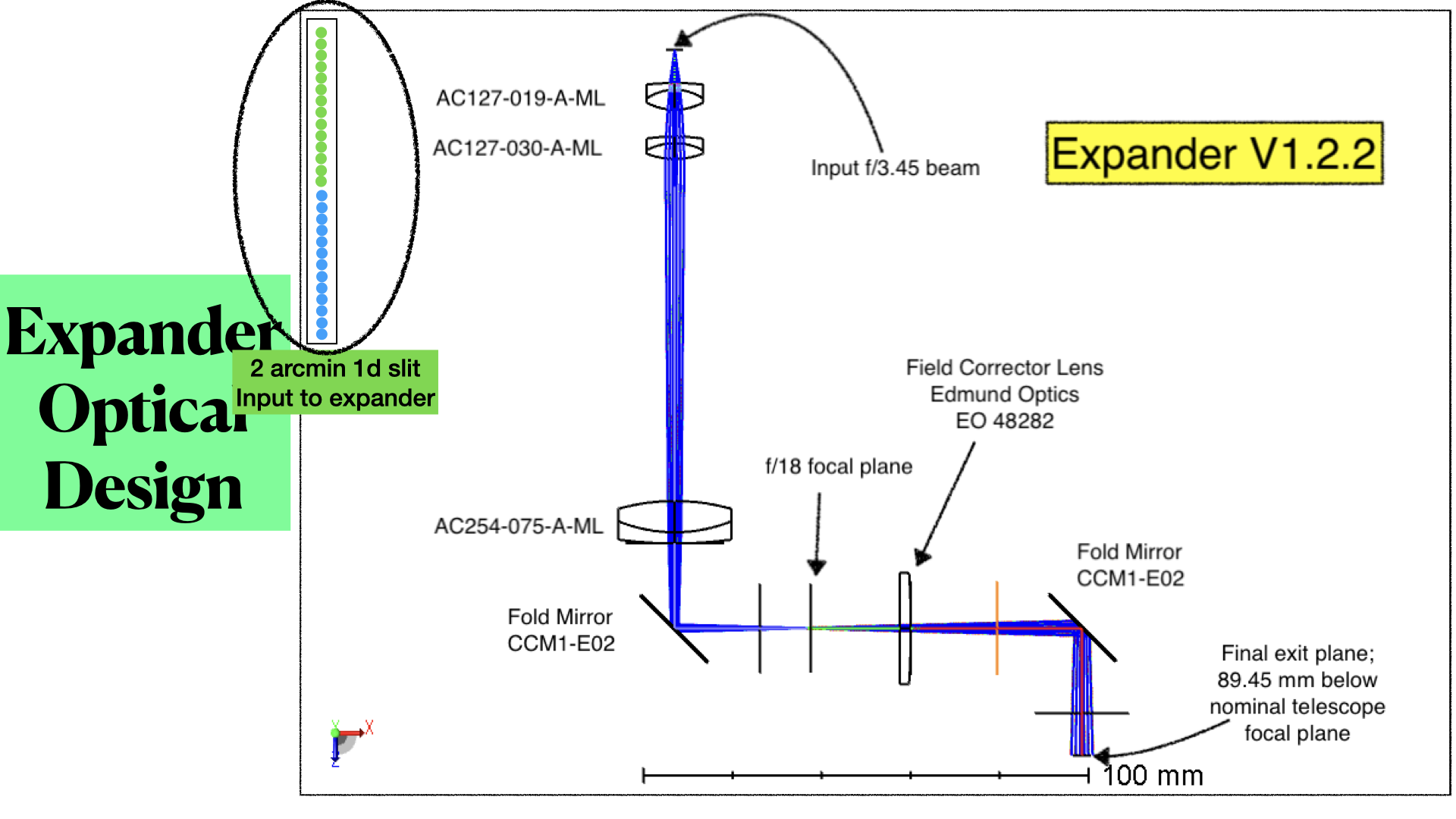}
    \caption{Optical layout of the ``Expander" subsystem takes as input the fiber pseudo-slit $f/3.45$ and creates a slow beam that matches the telescope beam at the plane on which FiberPol will be mounted on SpUpNIC. }
    \label{expander_optical_design}
\end{figure}

As noted in the previous sections, the ``Expander" subsystem takes as input the fiber pseudo-slit at $f/3.45$ and create an $f/18$ beam which matches the telescope beam at the slit input. Figure~\ref{expander_optical_design} shows the optical layout of the ``Expander" system. In addition to these optics, there is field lens that matches the field lens lying immediately after the SpUpNIC slit. This is required because in order to mount FiberPol on SpUpNIC, the existing slit mechanism is replaced whenever FiberPol is in use. The slit assembly of SpUpNIC consists of the pneumatic-slit as well as the field lens. Like in the ``Reducer'', multiple fold-mirrors are included to package the optical system and accurately mount the system on  the SpUpNIC focal plane assembly.

\subsection{Optomechanical Design}
\begin{figure}[h!]
    \centering
    \includegraphics[scale=0.25]{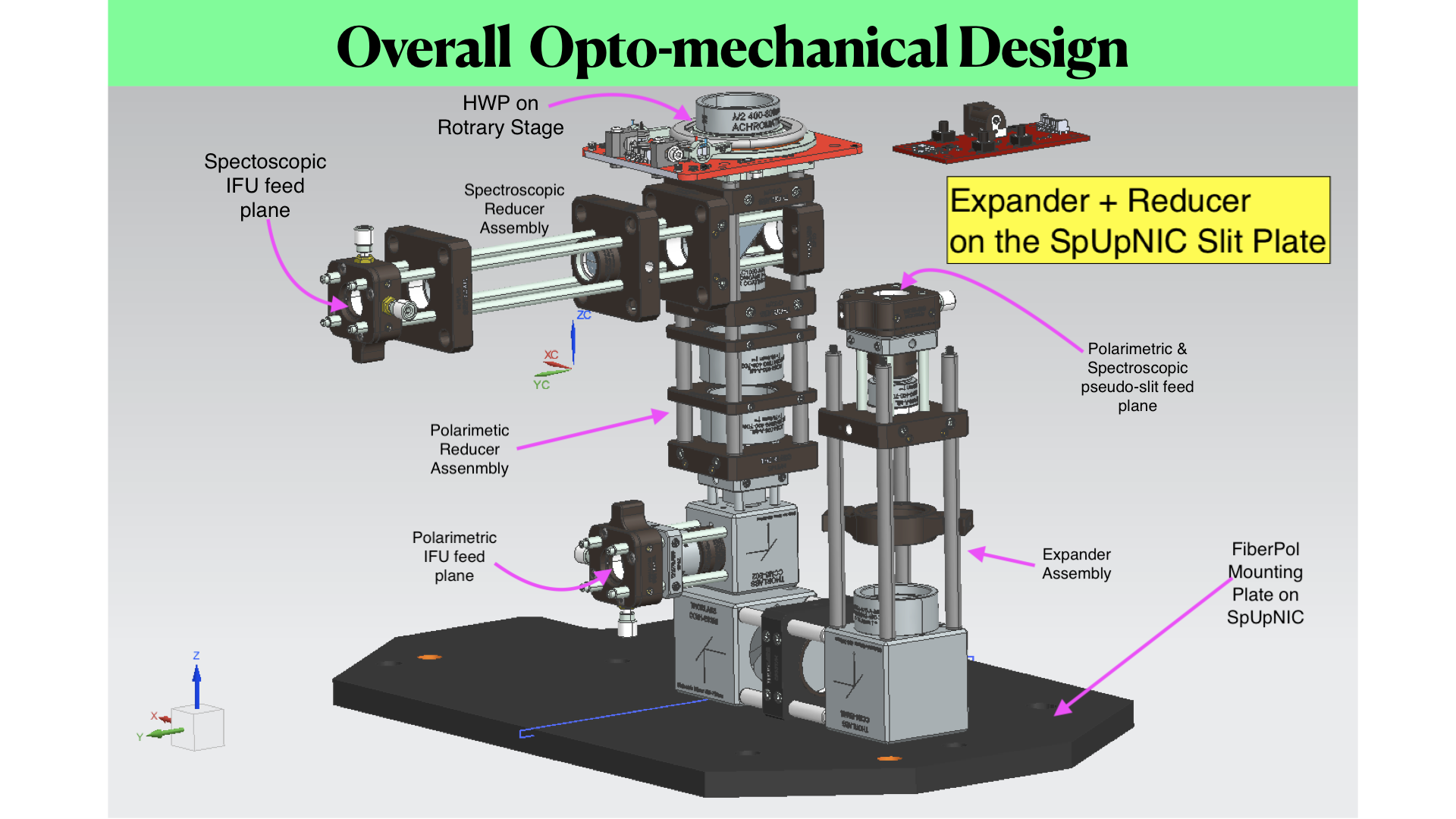}
    \caption{\footnotesize{Overall FiberPol design showing the entire optical system mounted inside the optomechanical system. Designed to be compact and a low-cost technology, it uses off-the-shelve low cost components, yet, delivers high quality (seeing limited) imaging and throughput ($\sim80\%$) performance. Its total height is under 20 cm. }}
    \label{FiberPol_design}
\end{figure}

Figure~\ref{FiberPol_design} shows the complete FiberPol model with all the optics mounted inside a cage plate system. The decision to design a cage-system kind optomechanical assembly was motivated by the following considerations: (a) we expect to meet the required alignment accuracies for the system using a cage-system, (b) they are easy to assemble, align and characterize in the lab, and (c) a large and diverse range of optomechanical mounts and stages (eg. rotary stage) compatible with a cage system are available in catalogs of vendors. As noted earlier, all mounts and components in the model are commercially available off-the-shelve items. Preliminary results from the testing of optical system presented in Section~\ref{colculsions_sec} indicate that optical system has been aligned to the required accuracy using the cage-system assembly. 

\begin{figure}
    \centering
    \includegraphics[scale = .24]{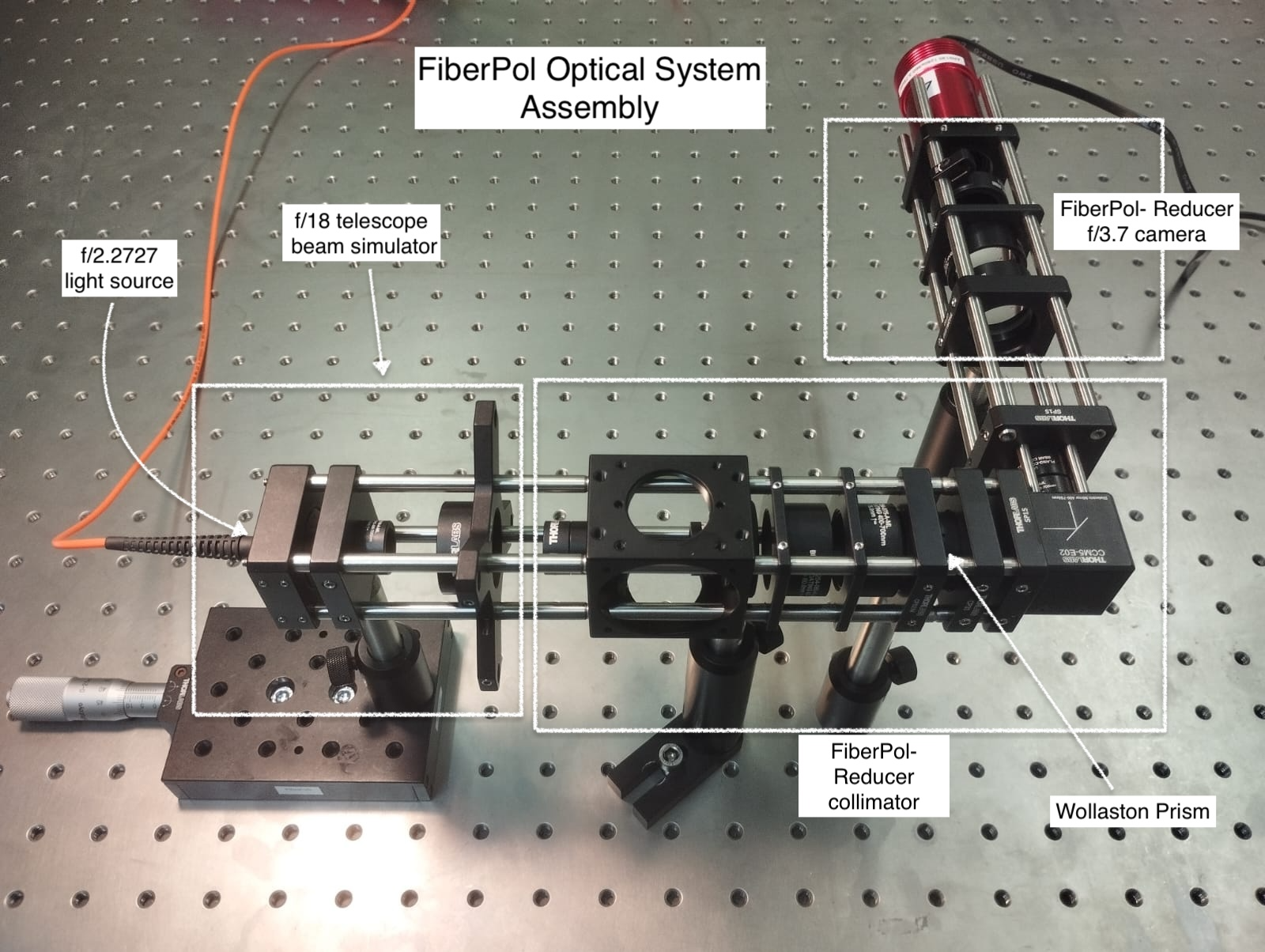}
    \caption{FiberPol's Reducer system assembled in the lab. A lab simulated $f/18$ beam, replicating the telescope beam is fed the system. This test-bed is used characterize the optical performance of the system.}
    \label{lab_assembly_model}
\end{figure}

\section{Current Status}
\label{colculsions_sec}

Figure~\ref{lab_assembly_model} shows the assembled ``Reducer" system of FiberPol and fed by a $f/18$ beam to emulate the beam coming from the telescope. The only difference to the final ``Reducer" assembly is the addition of a image relay system consisting of identical 75 mm focal length achromatic lens pair to allow sufficient gap between the last lens surface and the detector camera. The first test done on the system was to ascertain the separation of the $e$ and $o$ images at the focal plane. The nominal separation obtained from the optical design in 0.536~mm, while the measured separation value is $0.536\pm0.008$~mm, confirming that the FiberPol assembly in the lab is behaving as per design expectations.

\par All the optical and optomechanical components for the instrument have been procured. The instrument assembly and testing has commenced, with a scheduled on-sky commissioning for the later part of 2024. During the commissioning, we will demonstrate the technical capabilities of FiberPol and also carry out scientific observations, including the study of the ISM of nearby galaxies.

The development of an affordable and general purpose FiberPol like IFSP system will be a major advancement in the field of optical polarimetry instrumentation. It will have two lasting benefits to the astronomy community: 
\begin{enumerate}
    \item \textit{As a novel observational capability}, it will open up a new phase-space of observational  measurements i.e., spectropolarimetry of extended objects. Among the many scientific studies that this will enable, we will study the ISM of nearby galaxies through spectro-polarimetry. The objective of the study is to obtain the Serkowski curves ($p$ vs $\lambda$) and test the models of grain alignment as a function of galaxy type and galactic environment. Comets, asteroids and nebulae are also well suited to be studied through FiberPol.

\item \textit{Enabling (affordable) polarimetry on large (ELT) telescopes:} FiberPol can be modified for use on any existing spectrograph, especially on bigger telescopes like the 10 m SALT and the upcoming 30 m class telescopes. For example, applications to other existing spectrographs such as NIRWALS, HRS and RSS on the SALT could be effected with similar small-sized and low-cost optics. Polarimetric instrumentation is most often discarded  on large telescopes due to the requirements of large, expensive and fragile optics, and instrument development challenges. As an illustration,  none of the current first generation instruments for the three ELTs have polarimetric capabilities, even though the scientific merit for polarimetry with these instruments is generally agreed as compelling by the community. At the same time, FiberPol will make spectropolarimetry and IFSP technology accessible to existing small telescopes and observatories as they often cannot afford a dedicated polarimeter. 

\par Here we discuss briefly how to apply FiberPol on two potential telescopes: the 10~m SALT and the 25~m Giant Magellan Telescope (GMT)\cite{GMT}, which have direct Cassegrain-type ports available for instrument mounting and do not have Naysmyth-type tertiary mirror. SALT has a 10~m primary mirror and an $\rm f/4.2$ prime-focus plane\cite{salt_comissioning} that corresponds to a plate scale of $\sim$4.7"/mm. Using a similar optical system as FiberPol, we can employ a 25~mm lens immediately after the prime-focus plane\footnote{the first lens can be within 5-10 mm to minimize vignetting due to the expanding beam cone after the focus. If the first lens is placed farther, it will lead to a smaller FoV with same optics size or will need larger optics. Current catalog of off-the-shelve lenses have maximum apertures of 50~mm.} which will allow access to a field of view of 2 arcminutes. Unlike the current version of FiberPol, we will not need a field reducer lens to create an intermediate focus, and the collimator will be the first optical system. Further simplifying the implementation, there will be no need for an ``Expander" optical system: the fiber can be fed at a slightly slower $f/\#$ (using Equation~\ref{frd}) to compensate for FRD and match the telescope's $f/\#$ at the exit. In larger telescopes like SALT, we may need to increase the size of pupil and thus lens aperture used fed to the WPs in order to allow for sufficient separation between the $e$ and $o$ images at the fiber-injection plane.

\par The 25.4 m GMT, with its Gregorian-type design, produces an $f/8.2$ beam (1”/mm plate scale) at the focal plane after the wide-field corrector. A similar optical system to that described for SALT can be used, in addition to an Expander to match the telescope’s $f\#$ at the fiber exit plane. A 25 mm aperture will allow for a field of view of 20 arcseconds.
\par For telescopes with Nasmyth focal planes, implementing polarimetry requires additional optical and optomechanical elements (e.g., a derotator system) to measure and compensate for the polarization and other systematics introduced by the tertiary mirror, which changes in real-time during an observation.
\end{enumerate}



\acknowledgments 
 
The authors acknowledge the support of the National Research Foundation of South Africa for this project through the Salt Research Chair grant SARChI-114555.

\bibliography{article.bib} 

\begin{thebibliography}{10}

\bibitem{Hough_review}
Hough, J., ``{Polarimetry: a powerful diagnostic tool in astronomy},'' {\em Astronomy And Geophysics}~{\bf 47},  3.31--3.35 (06 2006).

\bibitem{Scarrott-1991}
Scarrott, S., ``Optical polarization studies of astronomical objects,'' {\em Vistas in Astronomy}~{\bf 34},  163 -- 177 (1991).

\bibitem{hippi2}
Bailey, J., Cotton, D.~V., Kedziora-Chudczer, L., De~Horta, A., and Maybour, D., ``Hippi-2: A versatile high-precision polarimeter,'' {\em Publications of the Astronomical Society of Australia}~{\bf 37} (2020).

\bibitem{DIPOL2}
Piirola, V., Berdyugin, A., and Berdyugina, S., ``{DIPOL-2: a double image high precision polarimeter},'' in [{\em Ground-based and Airborne Instrumentation for Astronomy V}{\nolinebreak\hspace{0.1em}]},  Ramsay, S.~K., McLean, I.~S., and Takami, H., eds.,  {\bf 9147},  2719 -- 2727, International Society for Optics and Photonics, SPIE (2014).

\bibitem{robopol}
Ramaprakash, A.~N., Rajarshi, C.~V., Das, H.~K., Khodade, P., Modi, D., Panopoulou, G., Maharana, S., Blinov, D., Angelakis, E., Casadio, C., Fuhrmann, L., Hovatta, T., Kiehlmann, S., King, O.~G., Kylafis, N., Kougentakis, A., Kus, A., Mahabal, A., Marecki, A., Myserlis, I., Paterakis, G., Paleologou, E., Liodakis, I., Papadakis, I., Papamastorakis, I., Pavlidou, V., Pazderski, E., Pearson, T.~J., Readhead, A. C.~S., Reig, P., Słowikowska, A., Tassis, K., and Zensus, J.~A., ``{RoboPol: a four-channel optical imaging polarimeter},'' {\em Monthly Notices of the Royal Astronomical Society}~{\bf 485},  2355--2366 (02 2019).

\bibitem{agn_unification}
{Antonucci}, R.~R.~J. and {Miller}, J.~S., ``{Spectropolarimetry and the nature of NGC 1068.},'' {\em Astrophysical Journal}~{\bf 297},  621--632 (Oct. 1985).

\bibitem{bershady20093d}
Bershady, M.~A., ``3d spectroscopic instrumentation,'' (2009).

\bibitem{Snik_keller_IFSP}
{Rodenhuis}, M., {Snik}, F., {van Harten}, G., {Hoeijmakers}, J., and {Keller}, C.~U., ``{Five-dimensional optical instrumentation: combining polarimetry with time-resolved integral-field spectroscopy},'' in [{\em Polarization: Measurement, Analysis, and Remote Sensing XI}{\nolinebreak\hspace{0.1em}]},  {Chenault}, D.~B. and {Goldstein}, D.~H., eds., {\em Society of Photo-Optical Instrumentation Engineers (SPIE) Conference Series} {\bf 9099},  90990L (May 2014).

\bibitem{SpUpNIC}
Crause, L.~A., Gilbank, D., van Gend, C., Worters, H.~L., Sass, C., Kotze, E.~J., Potter, S., Sickafoose, A., Sefako, R., Southworth, J., Macri, L., Thorstensen, J., Galan, C., Skelton, P., Engelbrecht, C., Braker, I., Winkler, H., Pieńkowski, D., S{\"u}rgit, D., Erdem, A., and Burleigh, M., ``{SpUpNIC (Spectrograph Upgrade: Newly Improved Cassegrain): a versatile and efficient low- to medium-resolution, long-slit spectrograph on the South African Astronomical Observatory’s 1.9-m telescope},'' {\em Journal of Astronomical Telescopes, Instruments, and Systems}~{\bf 5}(2),  024007 (2019).

\bibitem{andersson_review}
Andersson, B.-G., Lazarian, A., and Vaillancourt, J.~E., ``Interstellar dust grain alignment,'' {\em Annual Review of Astronomy and Astrophysics}~{\bf 53}(1),  501--539 (2015).

\bibitem{Serkowski1975}
{Serkowski}, K., {Mathewson}, D.~S., and {Ford}, V.~L., ``{Wavelength dependence of interstellar polarization and ratio of total to selective extinction.},'' {\em The Astrophysical Journal}~{\bf 196},  261--290 (Feb. 1975).

\bibitem{Hensley_2023}
Hensley, B.~S. and Draine, B.~T., ``The astrodust+pah model: A unified description of the extinction, emission, and polarization from dust in the diffuse interstellar medium,'' {\em The Astrophysical Journal}~{\bf 948},  55 (may 2023).

\bibitem{HARPSPol}
{Piskunov}, N., {Snik}, F., {Dolgopolov}, A., {Kochukhov}, O., {Rodenhuis}, M., {Valenti}, J., {Jeffers}, S., {Makaganiuk}, V., {Johns-Krull}, C., {Stempels}, E., and {Keller}, C., ``{HARPSpol {\textemdash} The New Polarimetric Mode for HARPS},'' {\em The Messenger}~{\bf 143},  7--10 (Mar. 2011).

\bibitem{PEPSI}
{Strassmeier}, K.~G., {Ilyin}, I., {J{\"a}rvinen}, A., {Weber}, M., {Woche}, M., {Barnes}, S.~I., {Bauer}, S.~M., {Beckert}, E., {Bittner}, W., {Bredthauer}, R., {Carroll}, T.~A., {Denker}, C., {Dionies}, F., {DiVarano}, I., {D{\"o}scher}, D., {Fechner}, T., {Feuerstein}, D., {Granzer}, T., {Hahn}, T., {Harnisch}, G., {Hofmann}, A., {Lesser}, M., {Paschke}, J., {Pankratow}, S., {Plank}, V., {Pl{\"u}schke}, D., {Popow}, E., and {Sablowski}, D., ``{PEPSI: The high-resolution {\'e}chelle spectrograph and polarimeter for the Large Binocular Telescope},'' {\em Astronomische Nachrichten}~{\bf 336},  324 (May 2015).

\bibitem{Skalidis}
{Skalidis, R.}, {Panopoulou, G. V.}, {Tassis, K.}, {Pavlidou, V.}, {Blinov, D.}, {Komis, I.}, and {Liodakis, I.}, ``Local measurements of the mean interstellar polarization at high galactic latitudes,'' {\em A\&A}~{\bf 616},  A52 (2018).

\bibitem{RoboPol_standards}
{Blinov, D.}, {Maharana, S.}, {Bouzelou, F.}, {Casadio, C.}, {Gjerløw, E.}, {Jormanainen, J.}, {Kiehlmann, S.}, {Kypriotakis, J. A.}, {Liodakis, I.}, {Mandarakas, N.}, {Markopoulioti, L.}, {Panopoulou, G. V.}, {Pelgrims, V.}, {Pouliasi, A.}, {Romanopoulos, S.}, {Skalidis, R.}, {Anche, R. M.}, {Angelakis, E.}, {Antoniadis, J.}, {Medhi, B. J.}, {Hovatta, T.}, {Kus, A.}, {Kylafis, N.}, {Mahabal, A.}, {Myserlis, I.}, {Paleologou, E.}, {Papadakis, I.}, {Pavlidou, V.}, {Papamastorakis, I.}, {Pearson, T. J.}, {Potter, S. B.}, {Ramaprakash, A. N.}, {Readhead, A. C. S.}, {Reig, P.}, {Słowikowska, A.}, {Tassis, K.}, and {Zensus, J. A.}, ``The robopol sample of optical polarimetric standards,'' {\em A\&A}~{\bf 677},  A144 (2023).

\bibitem{WALOP_Calibration_paper}
Maharana, S., Anche, R.~M., Ramaprakash, A.~N., Joshi, B., Basyrov, A., Blinov, D., Casadio, C., Deka, K., Eriksen, H.~K., Ghosh, T., Gjerl{\o}w, E., Kypriotakis, J.~A., Kiehlmann, S., Mandarakas, N., Panopoulou, G.~V., Papadaki, K., Pavlidou, V., Pearson, T.~J., Pelgrims, V., Potter, S.~B., Readhead, A. C.~S., Skalidis, R., Svalheim, T.~L., Tassis, K., and Wehus, I.~K., ``{WALOP-South: a four-camera one-shot imaging polarimeter for PASIPHAE survey. Paper II – polarimetric modeling and calibration},'' {\em Journal of Astronomical Telescopes, Instruments, and Systems}~{\bf 8}(3),  038004 (2022).

\bibitem{sky_pol}
{Maharana}, S., {Kiehlmann}, S., {Blinov}, D., {Pelgrims}, V., {Pavlidou}, V., {Tassis}, K., {Kypriotakis}, J.~A., {Ramaprakash}, A.~N., {Anche}, R.~M., {Basyrov}, A., {Deka}, K., {Eriksen}, H.~K., {Ghosh}, T., {Gjerl{\o}w}, E., {Mandarakas}, N., {Ntormousi}, E., {Panopoulou}, G.~V., {Papadaki}, A., {Pearson}, T., {Potter}, S.~B., {Readhead}, A.~C.~S., {Skalidis}, R., and {Wehus}, I.~K., ``{Bright-Moon sky as a wide-field linear Polarimetric flat source for calibration},'' {\em A\&A}~{\bf 679},  A68 (Nov. 2023).

\bibitem{FRD_Sabyasachi_paper}
Chattopadhyay, S., Bershady, M.~A., Wolf, M.~J., and Smith, M.~P., ``{Optimum telescope focal ratios for microlens-to-fiber coupled integral field spectrographs},'' {\em Journal of Astronomical Telescopes, Instruments, and Systems}~{\bf 8}(2),  025001 (2022).

\bibitem{WALOP_South_Optical_Design_Paper}
Maharana, S., Kypriotakis, J.~A., Ramaprakash, A.~N., Rajarshi, C., Anche, R.~M., Shrish, S., Blinov, D., Eriksen, H.~K., Ghosh, T., Gjerløw, E., Mandarakas, N., Panopoulou, G.~V., Pavlidou, V., Pearson, T.~J., Pelgrims, V., Potter, S.~B., Readhead, A. C.~S., Skalidis, R., Tassis, K., and Wehus, I.~K., ``{WALOP-South: a four-camera one-shot imaging polarimeter for PASIPHAE survey. Paper I—optical design},'' {\em Journal of Astronomical Telescopes, Instruments, and Systems}~{\bf 7}(1),  1 -- 24 (2021).

\bibitem{moptop}
Shrestha, M., Steele, I.~A., Piascik, A.~S., Jermak, H., Smith, R.~J., and Copperwheat, C.~M., ``{Characterization of a dual-beam, dual-camera optical imaging polarimeter},'' {\em Monthly Notices of the Royal Astronomical Society}~{\bf 494},  4676--4686 (04 2020).

\bibitem{GMT}
{Fanson}, J., {Bernstein}, R., {Ashby}, D., {Bigelow}, B., {Brossus}, G., {Burgett}, W., {Demers}, R., {Fischer}, B., {Figueroa}, F., {Groark}, F., {Laskin}, R., {Millan-Gabet}, R., {Park}, S., {Pi}, M., {Turner}, R., and {Walls}, B., ``{Overview and status of the Giant Magellan Telescope project},'' in [{\em Ground-based and Airborne Telescopes IX}{\nolinebreak\hspace{0.1em}]},  {Marshall}, H.~K., {Spyromilio}, J., and {Usuda}, T., eds., {\em Society of Photo-Optical Instrumentation Engineers (SPIE) Conference Series} {\bf 12182},  121821C (Aug. 2022).

\bibitem{salt_comissioning}
{Buckley}, D.~A.~H., {Barnes}, S.~I., {Burgh}, E.~B., {Crawford}, S., {Cottrell}, P.~L., {Kniazev}, A., {Nordsieck}, K.~H., {O'Donoghue}, D., {Rangwala}, N., {S{\'a}nchez}, R.~Z., {Sharples}, R.~M., {Sheinis}, A.~I., {V{\"a}is{\"a}nen}, P., and {Williams}, T.~B., ``{Commissioning of the Southern African Large Telescopes (SALT) first-generation instruments},'' in [{\em Ground-based and Airborne Instrumentation for Astronomy II}{\nolinebreak\hspace{0.1em}]},  {McLean}, I.~S. and {Casali}, M.~M., eds., {\em Society of Photo-Optical Instrumentation Engineers (SPIE) Conference Series} {\bf 7014},  701407 (July 2008).

\end{thebibliography}
\bibliographystyle{spiebib} 

\end{document}